%% file: main.tex
\documentclass[10pt,sigconf,letterpaper,nonacm]{acmart}

\usepackage{enumitem}
\usepackage{comment}
\usepackage{longtable}
\usepackage{multirow}
\usepackage{adjustbox}
\usepackage{natbib}
\usepackage{tcolorbox}
\usepackage{pgfplots}
\usepackage{graphicx}
\usepackage{booktabs}
\usepackage{textgreek}
\usepackage{listings}
\pgfplotsset{compat=1.18}

\usepackage{float}
\usepackage{url}

\usepackage[font=scriptsize]{subcaption} 
\usepackage{outlines}
\usepackage{enumitem}
\usepackage{multicol} 
\usepackage{placeins} 
\usepackage{tabularx}
\usepackage{wrapfig}

\usepackage{url}

\usepackage{booktabs}
\usepackage{siunitx}
\usepackage{tikz}

\usepackage[table]{xcolor}
\usepackage{colortbl}

\definecolor{low}{RGB}{198,239,206}      
\definecolor{medium}{RGB}{255,235,156}   
\definecolor{high}{RGB}{255,199,206}     

\usepackage{caption}

\usepackage{breakurl}

\usepackage[table]{xcolor}

\excludecomment{fullpaper}

\input{body/macros}

\begin{document}


\title{
Expecting (Targeted Ads)? Network Analysis of User Health Data Leakage in Fertility Tracking Apps
}

\author{\Large Yeeun Jo, Shahanaasree Sivakumar, Mahnoor Jameel, Camille Cobb, Adam Bates}
\email{{yeeunjo2, ss290, mjameel2, batesa, camillec}@illinois.edu}
\affiliation{
	\institution{University of Illinois at Urbana-Champaign}
	\streetaddress{610 E. John Street}
	\city{Champaign}
	\state{Illinois}
	\country{USA}
	\postcode{61820}	
}

\author{\Large Brad Reaves\vspace{5mm}}
\email{bgreaves@ncsu.edu}
\affiliation{
	\institution{North Carolina Statae University}
	\streetaddress{2101 Hillsborough Street}
	\city{Raleigh}
	\state{North Carolina}
	\country{USA}
	\postcode{27685}	
}

\begin{abstract}
\input{body/abstract}
\end{abstract}

\maketitle


\section{Introduction}
\label{sec:introduction}

\input{body/intro}


\section{Data Collection}
\label{sec:methods}
\input{body/methods}

\input{body/results}


\section{Discussion \& Conclusion}
\label{sec:conclusion}
\input{body/conclusion}

\section*{Acknowledgements}
We thank James O'Claire (AppGoblin) for open-sourcing his tools,
  helping us configure the network profiling environment,
  and meeting with us to share his insights into the Android advertising ecosystem.
This work was generously funded, in part, by the SPLICE research program, supported by a collaborative award from the National Science Foundation (NSF) SaTC Frontiers program under award number 1955228. 

{
\footnotesize
\bibliographystyle{ACM-Reference-Format}
\bibliography{jo-periodtracking-references,bates-bib-master} 
}

\appendix
\input{body/appendix}

\end{document}

%% file: body/macros.tex
\newif\ifdraft
\drafttrue

\newcommand{\gobble}[1]{ -- }    

\ifdraft
    \newcommand{\cc}[1]{\textcolor{purple}{[CC] #1}} 
    \newcommand{\yeeun}[1]{\textcolor{orange}{[Yeeun] #1}}
     
    \newcommand{\ab}[1]{\textcolor{olive}{[AB] #1}} 
    
    \newcommand{\redLogic}[1]{\textcolor{red}{\textit{#1}}}

\else
    \newcommand{\cc}[1]{}
    \newcommand{\yeeun}[1]{}
    
    \newcommand{\ab}[1]{}
    
    \newcommand{\redLogic}[1]{\textcolor{red}{\textit{}}}

\fi

%% file: body/abstract.tex
While human factors in the privacy of fertility tracking apps --
  health trackers that record users' menstrual or pregnancy data --
  has been the subject of extensive study,
  little attention has been paid to the technical aspects
  of apps' data handling practices.
We conduct a network-based measurement study of a corpus of
  20 Android fertility tracking apps from the Google Play Store, focusing on how
  user data is shared with third party advertising services.
After systematizing app features,
  we conduct a series of standardized user interactions across all apps
  in an environment that records TLS-stripped network traffic.
In a subset of apps (n=5) we identify explicit leakage of user health data
  as well implicit leakage through highly targeted contextual advertising URL's.
Equally importantly, we observe additional apps that use an ad-based monetization
  model without apparent leakage of user data,
  as well as several apps the interact only minimally with ad services.
These findings provide technical grounding for widespread user concerns,
  but also underscore the importance of consumer choice in the privacy
  implications of app-based fertility tracking.

%% file: body/intro.tex
Fertility tracking apps enable users to take daily notes and 
  track premenstrual syndrome, pregnancy symptoms, moods, intercourse, and period flow.
While these apps have continued to grow in popularity since their introduction,
  the highly sensitive and intimate information they collect has also been a long-held
  source of concern (e.g., \cite{chupaddados2017menstruappsh, rizk2016quantifying}).
In the United States,
  these concerns were further amplified following the Supreme Court's 2022 Dobbs decision,
  overturning Roe v. Wade (1973) and removing federal protections for abortion.

The digital privacy for fertility trackers is now a rich area of research,
  but prior work exhibits a noticeable methodological skew.
On the one hand, numerous studies have examined user's perceptions and mental models of
   fertility tracking privacy,
   conducting interviews \cite{mcdonald2023did},
   surveys \cite{cao2024deleted, hudig2025intimate},
   or examining social network posts \cite{song2024collective}
   to better understand how users reasoned about and managed privacy risks.
On the other hand,
   researchers have examined tracking vendors'
   public statements \cite{song2024our},
   privacy policies \cite{malki2024exploring, hassan2025unveilingprivacysecuritygaps}.
Studies oriented around users and policies constitute the majority of work in the space,
   providing valuable insights into both users' and vendors' attitudes and privacy management
   strategies in the ecosystem.

However, technical analysis of fertility tracking apps' actual data handling
  practices are underrepresented in the literature.
When software is addressed at all,
  prior work has typically conducted surface-level analysis as a means of grounding
  user- or policy-centered methods.
Malki et al. extend their policy analysis with a visual inspection of apps' user interfaces \cite{malki2024exploring}.        
Hudig et al. anchor their user survey by profiling the {\it encrypted} network traffic of 8 apps \cite{hudig2025intimate};
  however, with encrypted traffic they are unable to confirm or characterize the transmission of user data,
  instead establishing a correlation between active device usage and the volume of data transmission.
Hassan et al. analyze Android permissions and test apps for the presence of common third party trackers
  \cite{hassan2025unveilingprivacysecuritygaps}, again extending a policy analysis methodology.
Taken as a whole, prior studies of fertility trackers are limited in that they must
  speculate about how user privacy {\it might} be violated through app usage,
  grounding speculation through analysis of developer statements, privacy policies, access control lists, etc.
{\it To advance this area of study, what is required is technical ground truth --
  Do fertility trackers share user health data with third parties, and if so
  how fine-grained is the data being shared?}

As a first step to answering this question,
  we conduct a systematic analysis of fertility tracking apps'
  {\it unencrypted} network transmissions to third party advertising networks.
On a corpus of 20 apps,
  we conduct a series of 8 interaction sessions
  designed to exercise the common features of tracking apps.
These interactions were conducted in an emulated Android  environment instrumented
  with {\tt mitmproxy} to record TLS-stripped copies of network traffic.
After filtering the traffic captures for ad network domains,
  we catalog the role and function of ad network API endpoints
  then manually review URL query parameters and HTML request bodies
  for evidence of user health data leakage.
All instances of data leakage were then confirmed through manual review
  of decompiled app source code.
We organize our findings through the following high-level research questions.

\newcommand{\rqone}[0]{How do fertility trackers interact with advertising services?}
\newcommand{\rqtwo}[0]{Does advertising network traffic contain user health data?}

{\bf RQ1: \rqone} We identify interactions with 22 advertising network services and characterize their role and prominence
  within the fertility tracking ecosystem.
We observe that pregnancy journaling is associated with {\it at least 2.5 times}
  the volume of ad network activity when compared to other journaling features.
We go on to characterize this activity by the specific role of API Endpoints,
  demonstrating a broad continuum of ad-based monetization strategies.

{\bf RQ2: \rqtwo} We analyze 108,534 URL query parameters and 50,088 key-value pairs from HTML request bodies
  from our 7,829 HTTP request dataset, searching for evidence of data leakage.
In total, we find that 5 of 20 apps (25\%) leak sufficient data for ad network's
  to infer the user's high-level interaction patterns.
In 2 apps, we find that finer-grained user health data is leaked,
  including the user's pregnancy and current trimester,
  the age in weeks of the user's child postpartum,
  and the occurrence and approximate date of pregnancy loss.
We verify that these parameters were indeed populated with user-provided data
  through manual review of decompiled app source code.
These patterns are observed in extremely popular apps including
  {\it Baby Center}, {\it What To Expect}, {\it The Bump},
  which on Google Play reported 10M+, 5M+, and 1M+ installs respectively at the time of our analysis.

We conclude by considering the implications of our results.
While in some cases highly sensitive user health data is leaked to
  ad networks,
  our findings also demonstrate a spectrum of strategies for
  managing privacy and monetization in the fertility tracking ecosystem.
Notably, many apps serve ads without apparent transmission of user health data 
  to ad networks, while some apps are observed to forego ads altogether.
Provided that they are informed, this variance suggests
  the potential for users to engage in meaningful
  decision making based on their privacy concerns and management strategies.

%% file: body/methods.tex
\begin{fullpaper}
\input{tables/tab_mobile_apps_developers}
\end{fullpaper}

To explore the interconnection between advertising and fertility privacy, 
  we analyzed the network traffic emitted by tracking apps during normal user interactions.
Network traces were collected in October 2025 at a Midwest city in United States, 
  from a representative selection of Android apps
  from the Google Play store. 
The network traces were generated by a human experimenter
  entering scripted series of inputs
  across all apps in eight interaction sessions. 
Once the data was collected, 
  we developed tooling to identify active ad networks in each app, 
  recovered the purpose and function of different Ad Network API endpoints, 
  and reviewed URL and JSON data sent by the app to the network. 
We then compare our findings across apps. In this section, we describe how we selected apps, 
  structured the system side for network data profiling, 
  and conducted ad network activity profiling. 

\subsection{App Selection}

We primarily identified fertility apps through keywords searches
  on the Google Play Store using the following phrases:
  ``Period Tracker,'' ``Period Cycle'',  ``Fertility Tracker'' ``Ovulation Tracker,'' and ``Pregnancy Tracker \& Calendar.''
Our searches identified 
  254 total apps.
We then considered the app for inclusion in the study based on the following criteria:

\begin{itemize}[leftmargin=4mm]
\item Provides features related to menstruation, fertility, pregnancy, or menopause.
\item Accepts as input user data related to (in)fertility (e.g., period cycles, pregnancy, symptoms).
\item Includes calendar-based features, a proxy for self-tracking functionality.
\item Is a live service that is actively maintained.
\end{itemize}

We applied these criteria to filter from our consideration
  apps that did not handle sensitive user data related to fertility,
  and thus do not have the same privacy concerns as fertility trackers.
In addition to Google Play search results,
  we included 4 additional apps:
  1 app developed by a nonprofit organization (Euki); 
  1 app that appeared in a web advertisement served to one of the authors
    claiming to be a privacy-preserving period tracker (Stardust); 
  and 2 apps based on one of the authors' prior
experience with fertility trackers.

In all, 258 fertility tracking apps were in our candidate set, from which we selected 20 
        apps that represented a balanced mix of app categories, developer types, and
	install count estimates, ensuring coverage of both popular and niche
	applications.
Our corpus of apps is shown in Table \ref{tab:mobile_apps_developers} in the appendix.\footnote{Several additional apps with 1M+ downloads (Flo, Clue, Ovia Cycle, and FF App) were initially selected but were later excluded because they required single-sign-on authentication that did not work in our analysis environment, making them unusable.}
The Google Play reported number of installs in our sample ranged between $50$ and $100,000,000$,
  	while the average user rating was 4.4.
17 apps used free-with-ads monetization models,
  1 app exclusively used in-app-purchases,
  while 2 used a ``freemium'' model that supported either.

\input{tables/tab_interaction_sessions}

\subsection{Interaction Procedure}

Interaction sessions were designed around the core user features of fertility tracking apps
that were collecting user data. 
We began by conducting affinity diagraming 
to organize the fertility tracking apps' features into natural clusters based on shared characteristics
and emergent themes. 
This collaborative grouping process naturally revealed a hierarchical taxonomy, 
which we then formalized to categorize the apps' core functionalities.
Leveraging this taxonomy, we designed structured interaction sessions around representative user personas. 
For example, interaction sessions involved sequential inputs such as logging menstrual cycle start dates, recording symptoms (e.g., period flows or cramps), and generating ovulation predictions. 
These actions simulated user behavior to provoke ad network requests,
  which may contain sensitive health data (e.g., cycle phase indicators) alongside device identifiers. 
By specifying an interaction procedure for each
fertility tracking feature, the network traces become easier
to interpret and compare across apps. This also allows us to
control the information provided to the app and test for its
impact on the resulting ad network traffic.
Resulted interaction sessions are summarized in Table~\ref{tab:interaction_sessions}.

\subsection{Network Profiling Apparatus}

For each app, we download the Android APK files,
  then installed them in Android 11 
  in the Waydroid Emulator 1.5.1. 
Waydroid was set to default configurations,
  except that the environment was modified to
  intercept HTTPS traffic, strip TLS, and record the packet captures to disk.
Specifically, we used a rootkit to escalate privilege in the operating system,
  installed a custom self-signed Certificate Authority (CA) certificate
  to Android's certificate store,
  then deployed {\tt mitmproxy} with the new CA keypair.\footnote{Details at  \url{https://github.com/ddxv/mobile-network-traffic}.}

As this process was difficult to fully automate,
  we re-used the same Waydroid environment for all apps.
Unfortunately, Waydroid does not offer a snapshot
  feature, which would have allowed us to quickly
  reset the environment to a clean state for each app.
Instead, to minimize the risk of interference,
we force-stopped all apps in the Android settings prior to interacting
          with the application of interest.
Upon review of the {\tt mitmproxy} logs, we also confirmed
  that all network traces originated from the application of interest
  and not from other senders on the device.
  
\begin{figure}[t]
\includegraphics[width=\linewidth]{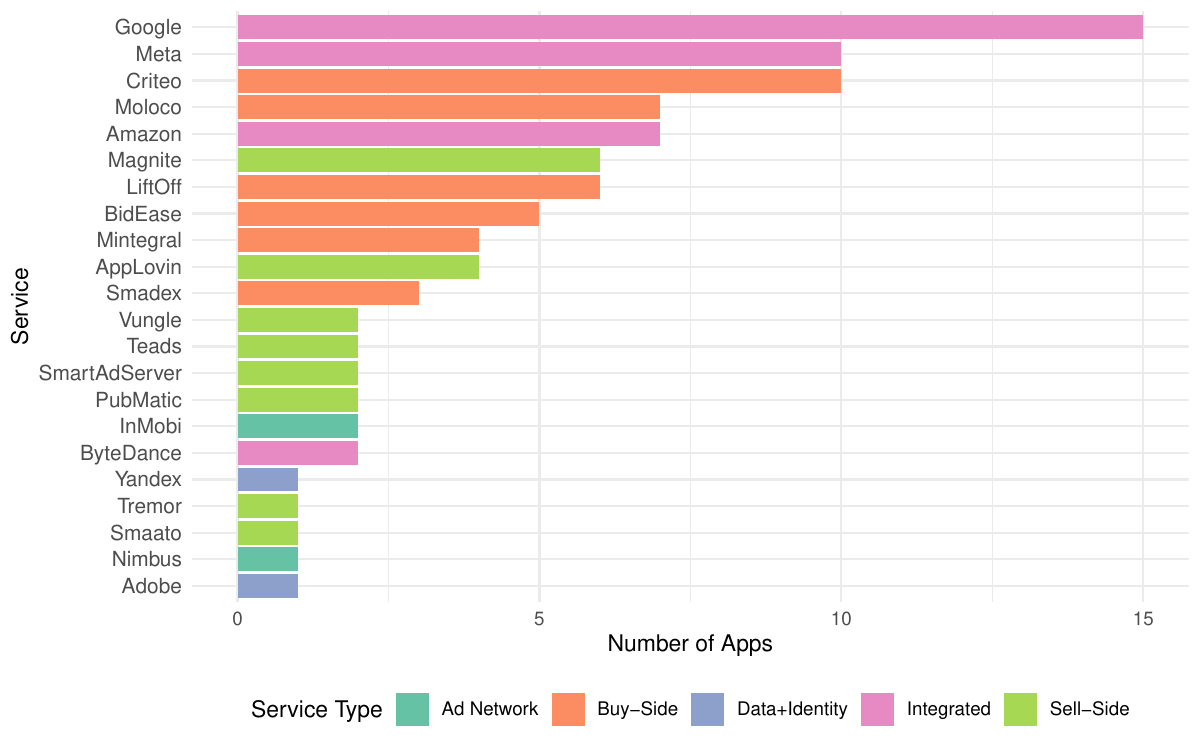}
\caption{Number of occurrences of each advertising service in the dataset.}
\label{fig:ad_network_app_counts}
\end{figure}

\subsection{Manual Code Review}

To validate and deepen our understanding of the advertising- related behaviors observed at the network level,
  we decompiled codes of each fertility tracking app.
 Using JADX decompiler, we decompiled the APKs and exported the reconstructed source code as Java files for comprehensive manual review.
 This process revealed direct correlations between network payloads and specific code paths. 
 In cases where we detect sensitive data transmission to ad networks -- such as fertility metrics, user demographics, or device identifiers -- manual code was used to verify that these values were dynamically constructed based on user data. 
 This dual network-static validation approach eliminated false positives and provided actionable evidence 
 of user data leakage patterns across all analyzed apps, strengthening the causal linkage between app 
 features and advertising ecosystem interactions.

%% file: tables/tab_mobile_apps_developers.tex
\begin{table*}[h]
\resizebox{\textwidth}{!}{
  \begin{tabular}{|l|l|l|l|l|l|l|}
  \hline
  \textbf{App Title} & \textbf{App ID} & \textbf{App Modes} & \textbf{Revenue Model} & \textbf{Installs} & \textbf{Dev. Name} & {\textbf{Headquarters}} \\ \hline
Clover & com.wachanga.womancalendar & Hybrid & FREE w/ ADS & 5M+ & Wachanga & {Limassol, Cyprus} \\
Stardust & com.stardust.app & Hybrid & FREE w/ ADS & 500K+ & Stardust App Inc. &  New York, NY, USA\\
My Calendar & com.popularapp.periodcalendar & Period & FREE w/ ADS & 100M+ & Simple Design Ltd & {Tortola, United Kingdom} \\
My Period Calendar & com.lbrc.PeriodCalendar & Period & IAP & 10M+ & SimpleInnovation &  Redmond, WA, USA\\
Lily Tracker & com.smsrobot.period & Period & FREE w/ ADS & 10M+ & SMSROBOT LTD & {Dublin, Ireland} \\
GP Tracker & com.period.tracker.lite & Period & FREEMIUM & 10M+ & GP International LLC &  Los Angeles, CA, USA\\
MIA & com.azarou.aventure.spoonge.run & Period & FREE w/ ADS & 1K+ & octave &  \\
BabyCenter & com.babycenter.pregnancytracker & Pregnancy & FREE w/ ADS & 10M+ & BabyCenter & {New York New York, USA} \\
Pregnancy + & com.hp.pregnancy.lite & Pregnancy & FREE w/ ADS & 50M+ & Philips Digital UK & {Noord-Holland-Amsterdam, Netherlands} \\
Amila Pregnancy & com.easymobs.pregnancy & Pregnancy & FREEMIUM & 5M+ & Amila & {Paphos, Cyprus} \\
What to Expect & com.wte.view & Pregnancy & FREE w/ ADS & 5M+ & What to Expect & {New York, NY, USA} \\
Pregnancy (duff hl studio) & com.women.pregnancytrackerapp22 & Pregnancy & FREE w/ ADS & 5M+ & duff hl studio &  \\
Preglife & com.gravid.gravid & Pregnancy & FREE w/ ADS & 1M+ & Preglife &  \\
The Bump & com.xogrp.thebump & Pregnancy & FREE w/ ADS & 1M+ & The Bump & {Chevy Chase, MD, USA} \\
Timskiy Pregnancy & com.timskiy.pregnancy & Pregnancy & FREE w/ ADS & 1M+ & Timskiy &  \\
Ovyu & com.ishitechnolabs.ovyu & Pregnancy & FREE w/ ADS & 1K+ & Ishi Technolabs & {Gujarat Surat, India} \\
Radhu Pregnancy & com.radhu.pregnancytracker & Pregnancy & FREE w/ ADS & 50+ & Radhu &  \\
Eveline & com.ixensor.lh & TTC & FREE w/ ADS & 100K+ & iXensor Co. Ltd. & {Taipei City, Taiwan} \\
Kindara & com.kindara.pgap & TTC & FREE w/ ADS & 100K+ & Kindara, Inc. & {Boulder, Colorado , USA} \\
Mira & mira.fertilitytracker.android\_us & TTC & FREE w/ ADS & 50K+ & Quanovate Tech Inc & {Pleasanton , CA, USA}\\ \bottomrule
  \end{tabular}
  }
\caption{Fertility Tracking Mobile Apps and Developer Company Headquarters. Revenue models: HARDPAY use required hardware purchase;
      FREE w/ ADS monetizes exclusively on ad revenue.
      IAP uses in-app purchases.
      FREEMIUM uses both ads and in-app purchases. 
      }
\label{tab:mobile_apps_developers}
\end{table*}

%% file: tables/tab_interaction_sessions.tex
\begin{table}[t!]  \centering
\scriptsize
\begin{tabular}{|p{2.75cm}|c|p{4cm}|}
\hline
\textbf{Interaction Session} & {\bf \# Apps} & \textbf{Data Types} \\
\hline

Registration & 13 & Personal Identification, (Reproductive State Indicators, Menstrual Activity ) \\ \hline

Menstruation Journaling & 16 & Menstrual Activity, Physical Symptoms, Emotional and Behavioral symptoms \\ \hline

Fertility Journaling & 14 & Hormonal or Cycle-specific Physiological Indicators, Sexual and Relationship Data \\ \hline

Sleep Journaling  & 15 & Lifestyle and General Health \\ \hline

Contraceptions Journaling  & 13 & Sexual and Relationship Data,  Physical Symptoms, Emotional and Behavioral symptoms, and Medical data \\ \hline

Pregnancy Journaling  & 15 & Sexual and Relationship Data, Reproductive State Indicators, Pregnancy Tracking, and Medical data \\ \hline

Postpartum Journaling  & 14 & Physical Symptoms, Emotional and Behavioral symptoms, and Medical Records \\ \hline

Pregnancy Loss Journaling & 15 & Reproductive State Indicators, Physical Symptoms, Emotional and Behavioral symptoms \\ \hline

\hline
\end{tabular}
\caption{Interaction Sessions used in Data Collection.}
\label{tab:interaction_sessions}
\end{table}

%% file: body/results.tex
`
\section{\rqone}

\begin{figure}
\includegraphics[width=\linewidth]{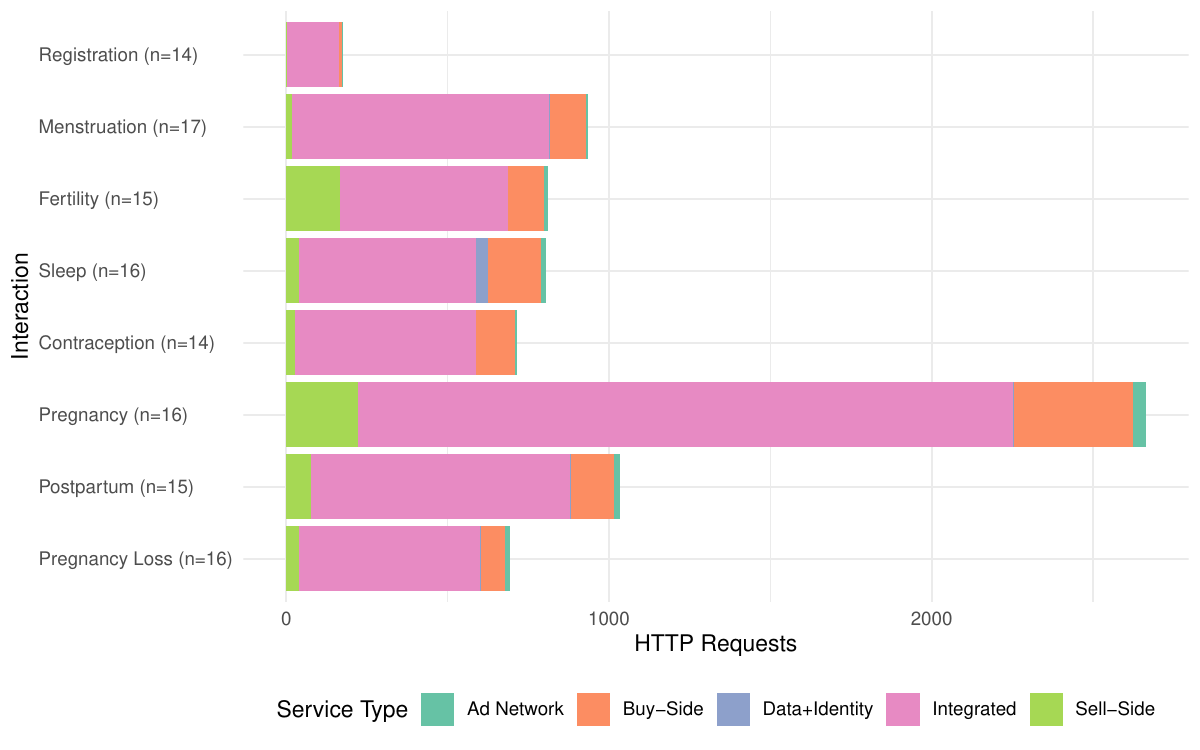}
\caption{Number of HTTP Requests by interaction session to different advertising network service types.}
\label{fig:requests_by_interaction_session}
\end{figure}

From the network traffic dataset, we filter
  for domains attributed to advertising services
  by {\tt appgoblin.info},\footnote{\url{https://github.com/appgoblin-dev/}}
  a site that monitors Android marketing and advertising analytics.

{\bf Service Type}
A high-level summary of observed services, by number of apps,
  can be found in Figure \ref{fig:ad_network_app_counts}.
We classify each service by its function in the ecosystem:
{\it Ad Networks} intermediaries that aggregate and resell ad inventory (space),
{\it Buy-Side} platforms help advertisers purchase ad space,
{\it Data and Identity} measurement platforms collect and unify consumer
                        information across different channels,
and {\it Sell-Side} services help developers to automatically sell ad space and mediate
                multiple ad networks competing.
Most prevalent in our data,
  vertically {\it Integrated} ad networks may perform all of these functions.

Figure \ref{fig:requests_by_interaction_session} reports how these different services
  contributed to network traffic in our dataset, grouped by interaction session.
Here, it can be seen that the vast majority of traffic
  is sent to the major integrated ad platforms, 
  which is consistent with the app-service peering relationships reported
  in Figure \ref{fig:ad_network_app_counts}.
We also observe large differences in the volume of ad traffic
  encountered in different interaction sessions --
  besides registration, most sessions ranged from 692 to 1,035 requests,
  but pregnancy interactions saw 2,663 requests.
App counts per interaction session,
  marked on the y-axis labels
  show this increase is not explained by the
  pregnancy interaction ($n=$15)
  being supported by more apps than other interactions ($n=$13-17).

\begin{figure}
\includegraphics[width=\linewidth]{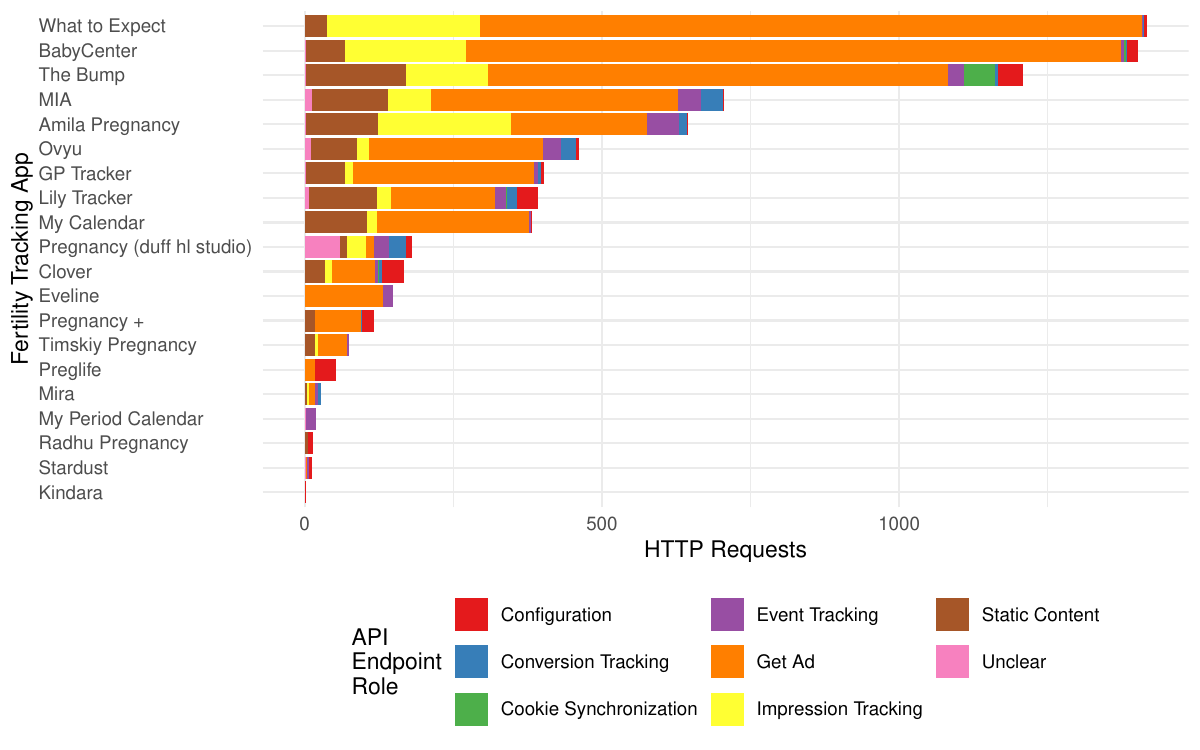}
\caption{Number of HTTP Requests by API Endpoint Role issued by different Fertility Trackers.}
\label{fig:requests_by_app}
\end{figure}

\begin{table}[t]
\footnotesize
\centering
\begin{tabular}{lrr}
\toprule
\textbf{Endpoint Role} & \textbf{Requests} & \textbf{Perc.} \\ \hline
Configuration          &   234 &  3\% \\ \hline 
Conversion Tracking    &   146 &  2\% \\ \hline 
Cookie Synchronization &    60 &  1\% \\ \hline 
Event Tracking         &   257 &  3\% \\ \hline 
Get Ad                 & 5,032 & 64\% \\ \hline 
Impression Tracking    & 1,022 & 13\% \\ \hline 
Static Content         &   979 & 13\% \\ \hline 
Unclear                &    99 & 1\% \\ 
\bottomrule
\end{tabular}
\caption{Request distribution by API Endpoint Role}
\label{tab:endpoint_roles}
\end{table}

\noindent
{\bf API Endpoint Roles.}
Large number of HTTP requests to ad services
  does not necessarily imply leakage of user halth data.
For example,many of these requests were for static content --
  no dynamic information is transmitted in the request.
To better characterize the privacy implications of
  this traffic,  
  we systematically categorized each of the 7,829 HTTP request by
  the function of the API Endpoint.  
To do so, we first defined the endpoints of an API as the set of
  paths associated with a domain that did not contain any dynamic
  or non-deterministic content.
Because dynamic content appeared later on the URL path,
  we define an API endpoint as a domain and path prefix pair.  
Of the resulting 196 endpoints, 
  we then reviewed public documentation and
  the request/response pairs in our dataset to
  determine their high-level function within the API.
Using this strategy we were able to identify a high-level role
  for 98.7\% of API Endpoints, the distribution for which
  are summarized in Table \ref{tab:endpoint_roles}.

\begin{figure}[t]
  \centering
    \includegraphics[width=.92\linewidth]{"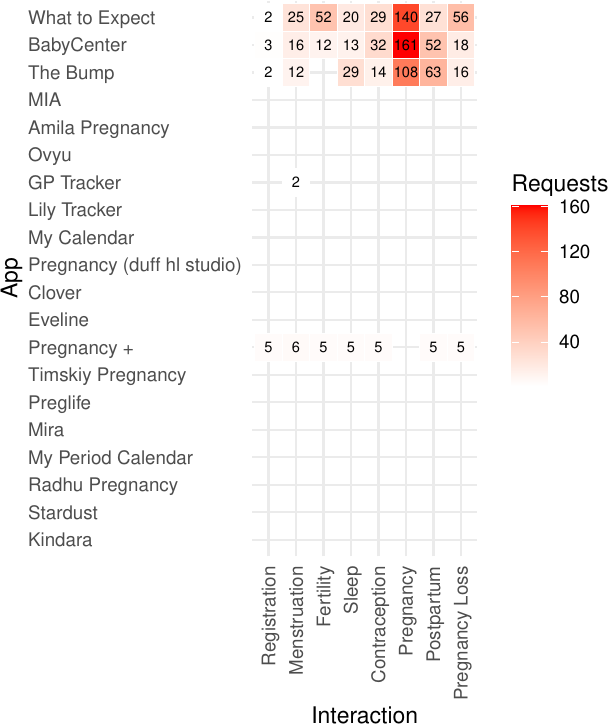"}
  \caption{Number of HTTP Requests per App and Interaction session that were labeled as leaking user data.}
  \label{fig:data_leakage}
\end{figure}

\section{\rqtwo}
\label{sec:rqtwo}

Having documented the widespread practice of ad-based monetization in the prior
  section, we continue by investigating whether this ad traffic
  contains user-provided health information as part of a targeted advertising strategy.
Specifically,
  we define {\it user data leakage}
  as any information that would permit a bystander to learn something about
  the user at a finer-granularity than the installation of the app itself.
As a guiding example from our data,
  a field containing the URL {\tt https://www.whattoexpect.com} is not data leakage,
  as this domain is associated with any use of {\it What to Expect};
  on the other hand,
  a field that references this domain with the URL path {\tt /pregnancy/after-miscarriage},
  it is data leakage because the bystander can infer that the user was
  journaling about a pregnancy loss.

Our procedure was as follows.
First, we split dynamic data from our 7,829 HTTP requests into
  key-value pairs,
  resulting in 
  108,534 tuples from URL query parameters
  and 50,088 from HTTP Request Body parameters.
Two experimenters then manually reviewed the contents of all fields
for evidence of data leakage.
The process of labeling all fields was made simpler by sorting by parameter key,
  allowing many tuples to be labeled after reviewing a small handful.
When key-value pairs contained complex or non-human-readable data,
  we attempted Base64 or URL decoding to interpret the value.    
We also consulted available documentation and community posts to
  interpret parameters that were not immediately interpretable.
Because of the opacity of ad network API's,
  as well as the potential for false negatives in our review procedure,   
  our analysis represents a conservative estimation of leakage in the dataset.
Inter-coder reliability and coder disagreement procedures were not needed,  
  as all explicit leakage of user data is verified in Appendix \ref{app:source_code}
  through review of decompiled app source code.

    \newcolumntype{L}{>{\raggedright\arraybackslash}X}

\begin{table}
\scriptsize
\begin{tabularx}{\linewidth}{|l|L|}
  \hline
  Interaction & Example Health Leakage Parameters\\ \hline
  Registration & {\tt us=}{\it precon}, {\tt us=}{\it preg02,tri1}\\ \hline
  Menstruation & {\tt appmode=}{\it ttc}, 
                 {\tt csw=}{\it preg02,tri1},
                 {\tt sid=}{\it pregnancy-prep-for-moms-to-be},
                 {\tt subcat=}{\it ovulation}\\ \hline
  Fertility & {\tt cat=}{\it gettingpregnant},
              {\tt subcat=}{\it preppingforpregnancy},
              {\tt contentidentifiers=}{\it conceptionbasics, preconceptiontests, naturalfamilyplanning
symptomsallkinds, vaginalbleedingdischarge}\\ \hline
  Sleep & {\it No unique values, repeats other leakage parameters} \\ \hline
  Contraception & {\tt topic=}{\it baby-products-health-and-safety}\\ \hline
  Pregnancy & {\tt bage=}{\it 33},
              {\tt cat=}{\it pregnancy},
              {\tt csw=}{\it preg02,tri1},
              {\tt sid=}{\it checklist-packing-a-hospital-bag, registry},
              {\tt spot=}{\it child-growth, isitsafe, birthprefs},
              {\tt subtopic=}{\it pregnancy-weight-gain},
              {\tt tid=}{\it weight-gain} \\ \hline
  Postpartum & {\tt appmode=}{\it baby},               
               {\tt cat=}{\it firstyear},
               {\tt csw=}{\it post7m,child0yr},
               {\tt phase=} {\it baby},
               {\tt sid=}{\it your-baby-week-1},
               {\tt subcat}={\it monthbymonth}\\ \hline
  Pregnancy Loss & {\tt subcat=}{\it pregnancyloss},
               {\tt wknum=}{\it 17}, {\tt us=}{\it null}\\ 
  \hline
\end{tabularx}
\caption{Example URL query parameters identified as leaking user health data.}
\label{tab:leakage_examples}
\end{table}

\subsection{Findings}
  
Our findings are summarized by App and Interaction session in Figure \ref{fig:data_leakage};
  this figure reports the number of HTTP requests that were found to leak some form of user data.
Examples of data leakage are found in Table \ref{tab:leakage_examples}.
Examining the specific conditions of data leakage,
  we identify several broad patterns, described below.

{\bf Explicit Leakage in Developer-Defined Parameters.}
In this pattern,
  which was observed in the Google services on 2 apps ({\it What to Expect}, {\it BabyCenter}),
  developers define custom parameters that explicitly correspond to user-provided health information.
Fields matching this pattern included {\tt us}, {\tt csw}, {\tt appmode},
  {\tt bage}, {\tt phase}, and {\tt wknum}.
In {\it BabyCenter},
  these fields tracked the stage of the pregnancy ({\tt csw=preg02,tri1})
  or the age of the newborn child ({\tt csw=post7m\_3w}),
  which was also possible in {\it What to Expect} ({\tt us=}{\tt preg17,tri2}, {\tt us=} {\tt post7m} {\tt \_3w,child0yr}).
Most concerningly, in {\it What to Expect} two parameters appear to
  explicitly signify pregnancy loss -- {\tt appmode} was set to {\tt NULL} only
  in the Pregnancy Loss session,
  while {\tt wknum} is set to {\tt 17} in this session but was {\tt NULL} in all other sessions,
  suggesting that the parameter is tracks the week of pregnancy loss.
To ensure that we did not misinterpret these fields,
  we reviewed the decompiled app source code that constructed these strings
  to confirm they are constructed from user data.

{\bf Implicit Leakage in Developer-Defined Parameters.}
While explicit leakage patterns were less common,
  we observed implicit leakage through Google's custom parameters in 5 apps.
Broadly, this data could be used to consistently infer how the
  user was interacting with the app, but typically nothing else.
Some parameters in this category stored extremely coarse-grained
  user health information, e.g., {\tt appmode=ttc} (trying to conceive) in Fertility journaling.
More often, these fields corresponded to developer-defined
  usage themes, e.g., {\it The Bump} defines a content carousel controlled
  by the {\tt sid} field with values such as {\tt pregnancy-prep-} {\tt for-moms-to-be}.
In the minimal leakage case, {\it Pregnancy +}
  statically set {\tt GAMAudience} to {\tt GAMAudience\_MaternalHealth} {\tt \_Emotional}
  in all interaction sessions, suggesting a partially implement
  targeted advertising strategy.
  
{\bf Explicit/Implicit Leakage in Contextual Ad URLs.}
We found that contextual ad URLs --
  which are intended to enable targeted advertising without sharing user information --
  regularly leaked user data in requests to Google and Amazon.
This pattern is observed in {\it What To Expect}, {\it BabyCenter}, {\it The Bump}, and {\it GP Tracker}, {\it Pregnancy +}.
Most often this leakage was implicit and could be used to infer the interaction session.
For example, the {\it BabyCenter} app links to {\tt babycenter.com}'s {\tt ovulation-calculator} in the fertility session, {\tt birth-class} in the Pregnancy Session, {\tt teaching\-your\-baby\-and\-toddler\-to\-sleep} in the Postpartum session, and {\tt grief-loss} in the Pregnancy Loss session.
In some cases, however, these contextual URL's were so scoped that they
  constituted explicit data leakage.
For example, contextual ads in {\it What to Expect} repeated the same fine-grained pregnancy information, e.g., {\tt month-by-month/month-7} in the Pregnancy session.

{\bf Implicit Leakage in Ad Inventory Heirarchies.}
We identify a final pattern in {\it The Bump},
  which implicitly leaks user interactions in the way they organized
  their available ad space.
Following an OpenRTB standard \cite{openrtb_gpid},
  ad requests to Amazon contain references to ad ``slots'' like {\tt /4879/Tri1.n\_} {\tt TB/tools\#m-top-input}
  where {\tt 4879} is a developer account ID,
  {\tt Tri1.n\_TB} is a collection of related content,
  {\tt tool} is a specific page or activity,
  and {\tt m-top-input} is a specific DOM element ID.
Requests for ads in the
  Pregnancy, Postpartum, and Menstruation  interactions
  reference the {\tt Tri[1,2,3].n\_TB},
  {\tt Postnatal.n\_TB}, and {\tt TTC.n\_TB} collections, respectively
However, this signal is less reliable than other implicit leakage due to
  pre-emptive fetching of many ads that are never displayed.  
For example, we see referenes to {\tt Tri1}, {\tt Tri2}, and {\tt Tri3}
  ad inventory even though our Pregnancy interaction described a first trimester pregnancy.
This suggests the app proactively fetches ads for many pages that are never visited,
  a common practice that is resolved through impression tracking and mediation mechanisms.

%% file: body/conclusion.tex
Confirming widely held concerns about the privacy implications of fertility tracking apps,
  we document two instances of user health data being explicitly transmitted to third party advertising services.
In addition, we identify several
  design patterns through which coarser-grained knowledge about users' app interactions is implicitly leaked.
While we are confident in the correctness of our results,
  the nature of our study design assures that our estimates of user health data leakage are conservative.
Several of the most popular fertility tracking apps were excluded from our study due to incompatibility
  with our HTTPS proxy.
Further, because our analysis depended on identifying human-interpretable strings in outbound HTTP traffic,
  any level of data encoding would lead to a false negative.
Some apps for which we did not find evidence of data leakage may have indeed be transmitting
  user health data without the presence of conspicuous substrings like ``preg,'' ``tri,'' ``ttc,'' etc.
Acknowledgment of this limitation underscores the overtness of the uncovered privacy violations.
Further, even the most na\"ive standardization of these custom parameters values would measurably improve
  user privacy, complicating attempts by third parties to make inferences about the fields' meanings.

Regardless of whether we identified all instances of data leakage in this app corpus,
  our findings sketch a richer continuum of monetization strategies in the fertility tracking ecosystem.
In addition to membership-based services that allow users to opt-out of ads,
  there appears to be a thriving plurality of apps that conduct ad-based monetization
  without extensive targeting based on user data.
Even among the 5 apps for which we find to engage in targeted advertising,
  important distinctions exist.
Some perform contextual targeting that implicitly leaks
  minimal data about user interactions,
  or organize their ad inventory in such a way that those interactions can be inferred.
In the most extreme cases,
  apps aggressively targeted based on highly-sensitive user health inputs.

The high degree of variance in the monetization strategies of fertility tracking apps
  should be viewed as an opportunity for user privacy.
Consumer choice -- which apps a user chooses to install -- could be associated
  with dramatically different privacy outcomes within this ecosystem.
Further, there appear to exist highly accessible ad-based trackers that generate
  revenue without user targeting.
The difficult task, of course, is communicating to users these data handling practices and the potential
  tradeoffs between apps, empowering them to make informed choices.
Our work is a first step towards documenting these practices and tradeoffs at a technical level;
  we are fortunate that fertility tracking is already a rich area of study in the usable privacy community,
  assuring there will be experts to investigate effective methods of translating this knowledge to users.
Additional steps in future work include approaching data leakage as a software analysis problem,
  which could be conducted using information flow analysis techniques (e.g., \cite{egc+2010}).
Our findings illuminate the breadth of implicit control flow patterns that would need to be accounted for
  in order for these approaches to be effective.

In this work, we have conducted 
  (to our knowledge) the first systematic study of how
  advertising informs the privacy of fertility tracking app usage.
Confirming widespread concerns that sensitive health data could be shared with third parties,
  we uncover several patterns through which user data can be explicitly or implicitly leaked
  to advertising networks.
Our network-based approach to the matter enabled us to quickly identify instances of
  data leakage, but could not holistically identify all sensitive data transmissions.
We thus view our work as a first step towards a more comprehensive understanding
  of privacy in the fertility tracking ecosystem.

%% file: body/appendix.tex
\section{Explicit Data Leakage Source Code Verification}
\label{app:source_code}

To verify that suspicious query parameters explicitly leaked user health data,
  we conducted manual review of decompiled app source code.
Foreknowledge of the specific (sub)strings of interest in the network data
  greatly simplified this review.
We omit detailed verification of the implicit leakage patterns for brevity
  and because they largely self-evident given their correlation with specific
  interaction sessions.

\subsection{BabyCenter}

The {\tt getAdsUserStage} found in {\tt com.babycenter.pregbaby.} {\tt api.model.ChildViewModel} is responsible
  for populating the {\tt csw} and {\tt us}.
As the APK retained function names, the
  significance of the lines of code are relatively easy to discern.
Line 5, 8, and 11 delineate stage building for the
  trying-to-conceive, pregnancy, and postpartum stages,
  respectively.
We can see in Line 11 that the postpartum stage is inferred
  by the stage name reaching 11 months (11m) and 3 weeks (3w)
  of pregnancy.
The strings constructed in lines 9 and 17 correspond to the values
  we see in the pregnancy and postpartum interaction sessions,
  respectively.

\begin{figure}[t]
\scriptsize
\input{lstings/bcgetadsuserstage.lsting}
\caption{In {\it BabyCenter}, the {\tt getAdsUserStage} function in {\tt com.babycenter.pregbaby.api.model.ChildViewModel} populates the {\tt csw} and {\tt us} custom parameters.}
\end{figure}

\subsection{What to Expect}

We are also able to easily identify the functions responsible for constructing suspicious query parameters in {\it What to Expect}.
Once again, the combination of string literals and function names provides strong evidence that
  these values are being constructed based on dynamic user inputs.

\begin{figure}[t]
\scriptsize
\input{lstings/wteadmanager.lsting}
\caption{In {\it What to Expect}, numerous functions in {\tt app/src/main/java/com/whattoexpect/ad/AdManager} support the construction of query parameters that leak user health data.}
\end{figure}

Notably absent from the {\tt AdManager} code is
  is explicit logic for constructing custom parameters related to pregnancy loss.
A more careful review of code suggests that this data leakage was inadvertent on the part
  of {\it What to Expect}.
We do not provide a code snippet here due to large volume of source code across multiple files,
  but describe the control flow path below:
{\tt AdManager} also contains a {\tt buildState} function that,
  among other things passes along the value that eventually initializes the {\tt wknum} field
  that signalled the pregnancy loss in our network analysis.
This function also contains a switch statement that 
  sets the value for {\tt appmode} by checking if a function argument is 2 (``no pregnancy''),
  3 (``healing''), 4 (``ttc''), 5 (``pregnancy''), 6 or 7 (``baby''),
  otherwise it defaults to {\tt NULL}.
We traced this input value back through the main app logic, whose functions lacked symbol names in the APK.
Here, a similar function checks the same variable but includes
  a case for 8 (``CHILD\_HARD\_STOP''), which we believe to be associated with the pregnancy loss journaling feature.
Setting case 8 to {\tt NULL} may have been an attempt to avoid transmitting
  pregnancy loss to the ad network.
However, since there appear to be few interaction paths where {\tt appmode=NULL},
  the pregnancy loss is implicitly leaked.
Further, back in {\tt buildState}, the string for {\tt wknum} is passed on
  regardless of the result of the switch statement.
As this value is (presumably) not updated again following a pregnancy loss,
  the approximate date of the pregnancy loss is passed to the ad network.

\input{tables/tab_mobile_apps_developers}
\section{Ethics}

{\bf Justification for Research.}
This work is motivated by ethical concerns surrounding
  the aggregation and sharing of fertility data.
Prior work has helped to document the widespread public concern regarding
  the privacy policies of fertility trackers,
  including the risk of data access by law enforcement (e.g., \cite{cao2024deleted}).
However, to our knowledge there has been no study that attempts to assess
  the prevalence of such data sharing on a technical level.
Due to recent confirmation of federal law enforcement agencies using
  real-time bidding markets as a means of sourcing surveillance data \cite{eff_lea_advertising},
  the user data that fertility tracking apps transmit to advertising services
  is highly relevant to this broader societal concern and thus worthy of study.

{\bf Stakeholder Analysis.} This ecosystem is comprised of three primary stakeholders:
  (1) {\it Users} of fertility tracking apps that generate health data,
  (2) {\it Developers} of those apps that determine how that data is managed and shared,
  and (3) {\it Advertising Services} that developers establish business relationships with
  in order to monetize the app.

{\bf Potential Impacts.}
Our work has potential positive impacts, specifically for users,
  who may become more informed about the privacy implications of using different apps,
  allowing them to choose apps whose data handling practices are more aligned with
  their privacy concerns.
Informed consumer choice could lead to positive incentive structures for apps
  to adopt more privacy-preserving data sharing practices.
This argument is consistent with the Mozilla Foundation's {\it Privacy Not Included}
  project, which includes recommendations on reproductive health apps \cite{mozilla_reproductive_health}.
We plan to share our findings with Mozilla and other privacy advocacy groups following submission.
We also plan to notify {\it What to Expect} of the apparent inadvertent disclosure of
  pregnancy loss in their {\tt AdManager} class.
Our work will also benefit the privacy research community through the open sourcing of our data
  and analysis code following publication.

On the other hand, our work may have potential negative impacts for
  some app developers.
To our knowledge our analysis did not violate the terms of service of these applications --
  we did not modify the app and installing HTTPS proxies is a common organizational
  security practice --  nonetheless developers may be surprised to discover
  we were able to review their HTTPS requests to ad networks.
For apps that were shown to engage in highly targeted advertising
  based on transmission of user health data, the developers may face financial consequences
  for the publication of this work, however unlikely.
We feel that in both cases the risks of negative impacts are outweighed
  by the potential public good of disseminating these findings,
  especially in the latter case where the negative impact to one developer
  may be offset by a positive impact to another developer that can offer
  users greater privacy.

{\bf Mitigations.}
Our study is observational in nature; it does not propose a new defense
  nor provide instructions for launching a new attack.
An attacker could potentially leverage knowledge of this documented data leakage
  to surveil a targeted user by conducting their own man-in-the-middle attack
  on the user's device.
However, we find this risk esoteric and unlikely, as an attacker in a position to
  launch such an attack would have more direct means of surveilling the user.
Therefore, we pursue no mitigation beyond disseminating our results
  with concerned parties and advocacy groups, which we plan to undertake following submission.

%% file: main.bbl
\begin{thebibliography}{10}
\providecommand{\url}[1]{#1}
\csname url@samestyle\endcsname
\providecommand{\newblock}{\relax}
\providecommand{\bibinfo}[2]{#2}
\providecommand{\BIBentrySTDinterwordspacing}{\spaceskip=0pt\relax}
\providecommand{\BIBentryALTinterwordstretchfactor}{4}
\providecommand{\BIBentryALTinterwordspacing}{\spaceskip=\fontdimen2\font plus
\BIBentryALTinterwordstretchfactor\fontdimen3\font minus
  \fontdimen4\font\relax}
\providecommand{\BIBforeignlanguage}[2]{{%
\expandafter\ifx\csname l@#1\endcsname\relax
\typeout{** WARNING: IEEEtran.bst: No hyphenation pattern has been}%
\typeout{** loaded for the language `#1'. Using the pattern for}%
\typeout{** the default language instead.}%
\else
\language=\csname l@#1\endcsname
\fi
#2}}
\providecommand{\BIBdecl}{\relax}
\BIBdecl

\bibitem{chupaddados2017menstruappsh}
N.~Felizi and J.~Varon, ``Menstruapps - how to turn your period into money (for
  others),''
  \url{https://chupadados.codingrights.org/en/menstruapps-como-transformar-sua-menstruacao-em-dinheiro-para-os-outros-2/},
  2017.

\bibitem{rizk2016quantifying}
V.~Rizk and D.~Othman, ``\BIBforeignlanguage{English}{Quantifying fertility and
  reproduction through mobile apps: a critical overview.}''
  \emph{\BIBforeignlanguage{English}{Arrow for Change}}, vol.~22, p. 13–21,
  2016.

\bibitem{mcdonald2023did}
N.~Mcdonald and N.~Andalibi, ````i did watch ‘the handmaid's tale'': Threat
  modeling privacy post-roe in the united states,'' \emph{ACM Transactions on
  Computer-Human Interaction}, vol.~30, no.~4, pp. 1--34, 2023.

\bibitem{cao2024deleted}
J.~Cao, H.~Laabadli, C.~H. Mathis, R.~D. Stern, and P.~Emami-Naeini, ````i
  deleted it after the overturn of roe v. wade'': Understanding women's privacy
  concerns toward period-tracking apps in the post roe v. wade era,'' in
  \emph{Proceedings of the 2024 CHI Conference on Human Factors in Computing
  Systems}, 2024, pp. 1--22.

\bibitem{hudig2025intimate}
A.~I. Hudig and J.~Singh, ``Intimate data sharing: Enhancing transparency and
  control in fertility tracking,'' in \emph{Proceedings of the 2025 CHI
  Conference on Human Factors in Computing Systems}, 2025, pp. 1--24.

\bibitem{song2024collective}
Q.~Song, R.~Ma, Y.~Kou, and X.~Gui, ``Collective privacy sensemaking on social
  media about period and fertility tracking post roe v. wade,''
  \emph{Proceedings of the ACM on human-computer interaction}, vol.~8, no.
  CSCW1, pp. 1--35, 2024.

\bibitem{song2024our}
Q.~Song, R.~H. Hernandez, Y.~Kou, and X.~Gui, ``“our users' privacy is
  paramount to us”: A discourse analysis of how period and fertility tracking
  app companies address the roe v wade overturn,'' in \emph{Proceedings of the
  2024 CHI Conference on Human Factors in Computing Systems}, 2024, pp. 1--21.

\bibitem{malki2024exploring}
L.~M. Malki, I.~Kaleva, D.~Patel, M.~Warner, and R.~Abu-Salma, ``Exploring
  privacy practices of female mhealth apps in a post-roe world,'' in
  \emph{Proceedings of the 2024 CHI Conference on Human Factors in Computing
  Systems}, 2024, pp. 1--24.

\bibitem{hassan2025unveilingprivacysecuritygaps}
\BIBentryALTinterwordspacing
M.~Hassan, M.~Jameel, T.~Wang, and M.~Bashir, ``Unveiling privacy and security
  gaps in female health apps,'' 2025. [Online]. Available:
  \url{https://arxiv.org/abs/2502.02749}
\BIBentrySTDinterwordspacing

\bibitem{openrtb_gpid}
\BIBentryALTinterwordspacing
G.~D. Documentation, ``{OpenRTB Integration},'' 2026. [Online]. Available:
  \url{https://developers.google.com/authorized-buyers/rtb/openrtb-guide}
\BIBentrySTDinterwordspacing

\bibitem{egc+2010}
W.~Enck, P.~Gilbert, B.-G. Chun, L.~P. Cox, J.~Jung, P.~McDaniel, and A.~N.
  Sheth, ``{TaintDroid: An Information-flow Tracking System for Realtime
  Privacy Monitoring on Smartphones},'' in \emph{Proceedings of the 9th USENIX
  Symposium on Operating Systems Design and Implementation}, ser. OSDI'10, Oct.
  2010.

\bibitem{eff_lea_advertising}
\BIBentryALTinterwordspacing
L.~Cohen and H.~Hongo, ``{The Government Uses Targeted Advertising to Track
  Your Location. Here's What We Need to Do.}'' 2026. [Online]. Available:
  \url{https://www.eff.org/deeplinks/2026/03/targeted-advertising-gives-your-location-government-just-ask-cbp}
\BIBentrySTDinterwordspacing

\bibitem{mozilla_reproductive_health}
\BIBentryALTinterwordspacing
{Mozilla Foundation}, ``{*Privacy Not Included: Reproductive Health},'' 2026.
  [Online]. Available:
  \url{https://www.mozillafoundation.org/en/privacynotincluded/categories/period-ovulation-trackers/}
\BIBentrySTDinterwordspacing

\end{thebibliography}
